\documentclass[journal=jacstat,manuscript=article,showpacs,10pt,longbibliography]{achemso}

\usepackage[version=3]{mhchem} 
\usepackage{graphicx} 
\usepackage{amsmath}
\usepackage{float}
\usepackage{algorithm}
\usepackage{algpseudocode}
\usepackage{bm}
\usepackage{threeparttable}
\usepackage{physics}
\usepackage{amssymb}
\usepackage{multicol}
\usepackage{listings}
\usepackage{soul}
\usepackage{relsize}
\usepackage{makecell}
\usepackage{placeins}
\usepackage{color,soul}
\usepackage{flexisym}
\usepackage{expl3}
\usepackage{booktabs}
\usepackage{braket}
\usepackage{amsmath}

\newcolumntype{K}[1]{>{\arraybackslash}p{#1}}
\author{Abhishek Mitra}
\affiliation[University of Chicago]
{Department of Chemistry, Chicago Center for Theoretical Chemistry, University of Chicago, Chicago, IL 60637, USA.}
\author{Ruhee D'Cunha}
\affiliation[University of Chicago]
{Department of Chemistry, Chicago Center for Theoretical Chemistry, University of Chicago, Chicago, IL 60637, USA.}
\author{Qiaohong Wang}
\affiliation[University of Chicago]
{Pritzker School of Molecular Engineering, Chicago Center for Theoretical Chemistry, University of Chicago, Chicago, IL 60637, USA.}
\author{Matthew R. Hermes}
\affiliation[University of Chicago]
{Department of Chemistry, Chicago Center for Theoretical Chemistry, University of Chicago, Chicago, IL 60637, USA.}
\author{Yuri Alexeev}
\affiliation{Argonne National Laboratory
9700 S. Cass Avenue
Lemont, IL 60439}
\author{Stephen K. Gray}
\email{gray@anl.gov}
\affiliation{Argonne National Laboratory
9700 S. Cass Avenue
Lemont, IL 60439}
\author{Matthew Otten}
\email{mjotten@wisc.edu}
\affiliation[HRL Laboratories]
{Department of Physics, University of Wisconsin -- Madison, Madison, WI 53726, USA.}
\author{Laura Gagliardi}
\email{lgagliardi@uchicago.edu}
\affiliation[University of Chicago]
{Department of Chemistry, Chicago Center for Theoretical Chemistry, University of Chicago, Chicago, IL 60637, USA.}
\alsoaffiliation[University of Chicago]
{Pritzker School of Molecular Engineering, Chicago Center for Theoretical Chemistry, University of Chicago, Chicago, IL 60637, USA.}
\alsoaffiliation{Argonne National Laboratory
9700 S. Cass Avenue
Lemont, IL 60439}
\title[An \textsf{achemso} demo]
  {The Localized Active Space Method with Unitary Selective Coupled Cluster}
\abbreviations{IR,NMR,UV}
\begin{document}
\begin{abstract}
We introduce a hybrid quantum-classical algorithm,
the localized active space unitary selective coupled cluster singles and doubles (LAS-USCCSD) method. 
Derived from the localized active space unitary coupled cluster (LAS-UCCSD) method, LAS-USCCSD first performs a classical LASSCF calculation, then selectively identifies the most important parameters (cluster amplitudes used to build the multireference UCC ansatz) for restoring inter-fragment interaction energy using this reduced set of parameters with the variational quantum eigensolver method.
We benchmark LAS-USCCSD against LAS-UCCSD by calculating the total energies of $(\mathrm{H}_2)_2$, $(\mathrm{H}_2)_4$ and \textit{trans}-butadiene, and the magnetic coupling constant for a bimetallic compound [Cr$_2$(OH)$_3$(NH$_3$)$_6$]$^{3+}$. For these systems, we find that LAS-USCCSD reduces the number of required parameters and thus the circuit depth by at least one order of magnitude, an aspect which is important for the practical implementation of multireference hybrid quantum-classical algorithms like LAS-UCCSD on near-term quantum computers.
\end{abstract}

\maketitle

\section{Introduction}

Accurately modeling strong electron correlation is crucial in quantum chemistry, especially for describing transition-metal and heavy-metal chemistry, magnetic molecules, bond breaking, biradicals, excited states, and the functionality of various materials. Despite its significance, addressing this challenge with computationally feasible electronic structure methods remains daunting. 
Strong electron correlation, also known as ``static" or ``non-dynamic" correlation, emerges from degenerate or near-degenerate electronic states within systems referred to as multiconfigurational.~\cite{Helgaker2000} 
To investigate these systems effectively, multireference methods are indispensable, such as the complete active space self-consistent field (CASSCF) method~\cite{CASSCF0,CASSCF1,CASSCF2}  and multireference configuration interaction (MRCI)~\cite{Buenker1974,Buenker1978}. While these methods offer accuracy, their applicability is constrained by exponential and factorial scaling with the size of the ``active space," determined by the number of electrons and orbitals subjected to exact diagonalization~\cite{Lowdin1955}.

To enhance the practicality of multireference methods, approximate classical CI solvers like the density matrix renormalization group (DMRG)~\cite{White1992,Chan2002}, selected configuration interaction methods~\cite{Stein2019,li2018fast,li2020accurate}, and quantum Monte Carlo methods ~\cite{ciapprox-1,ciapprox-2,Booth2009} have been developed. These methods aim to reduce the exponential computational cost by eliminating less significant configurations. Another approach involves active space fragmentation techniques, such as active space decomposition (ASD)~\cite{Parker2013,Parker2014-0,Parker2014-1,Kim2015}, cluster mean-field~\cite{Jimenez-Hoyos2015}, rank-one basis states~\cite{Nishio2019,Nishio2022}, TPSCI algorithm~\cite{Abraham2020}, density matrix embedding theory (DMET)\cite{DMET,DMETpractical,DMET_solid_Cui,pDMET-1D,mitra2021excited,Mitra2022,Mitra2023,haldar2023local,Verma2023}, restricted active space (RAS)~\cite{Malmqvist1990,Olsen1988}, generalized active space (GAS)\cite{Ma2011}, and the localized active space self-consistent field method (LASSCF)~\cite{LASSCF0,LASSCF1}, among others. LASSCF, in particular, models strong, localized correlation within specific fragments while modeling inter-fragment correlation using a mean-field approach. However, when inter-fragment electron correlation beyond the mean-field level becomes important, LASSCF becomes inaccurate, as demonstrated in the tris-(\(\mu\)-hydroxo)-bridged chromium compound ([Cr$_2$(OH)$_3$(NH$_3$)$_6$]$^{3+}$), where LASSCF incorrectly suggests a high-spin sextet ground state instead of the actual low-spin singlet~\cite{Otten2022}.

To restore the electron correlation between fragments, the localized active space state interaction (LASSI) method was developed~\cite{Pandharkar2022}, though it reintroduces the factorial scaling of CASSCF. Another standard approach that can be used to restore electron correlation between fragments is the unitary coupled cluster method (UCC)~\cite{Bartlett1989, Taube2006} for inter-fragment excitations. However, when this method is applied to a LAS reference wave function, and in general any multireference wave function, it necessitates arbitrary truncation of the otherwise non-terminating equations of the cluster expansion for practical implementation on a classical computer. This has inspired the development of the LAS-UCC method where the UCC step can in principle be efficiently implemented on a quantum circuit simulator or quantum computer.~\cite{2020}. 

There have been significant advancements in quantum computing hardware and algorithms, particularly in the field of quantum chemistry simulation~\cite{Kim2023,Yu2023,https://doi.org/10.48550/arxiv.2309.02863,https://doi.org/10.48550/arxiv.2308.04481,https://doi.org/10.48550/arxiv.2307.07552,PhysRevResearch.5.013183}. This progress includes the development of various ansatzes such as the UCC ansatz\cite{Lee2018,Hoffmann1988,McArdle2020}, hardware efficient ansatzes,\cite{Kandala2017} adaptive structure anzatzes,\cite{Grimsley2019} and qubit coupled-cluster ansatzes\cite{Tang2021,Ryabinkin2020,ryabinkin2018qubit} for variational quantum eigensolver (VQE) simulations\cite{Peruzzo2014}, which is currently the most practical hybrid quantum-classical method in quantum chemistry. 

The LAS-UCCSD algorithm, while not yet run on actual quantum hardware, has been validated using quantum circuit simulators on a classical computer. It has been demonstrated that LAS-UCCSD achieves chemical accuracy (1 kcal/mol) in calculating total energies and an accuracy of  1 cm$^{-1}$ in calculating magnetic coupling constants, relative to the corresponding CASCI values, for the limited systems explored\cite{Otten2022,Dcunha2023StatePreparation}. However, a LAS-UCCSD calculation of moderately large systems such as the tris-($\mu$-hydroxo)-bridged chromium compound [Cr$_2$(OH)$_3$(NH$_3$)$_6$]$^{3+}$ with a mixed def2-TZVP/def2-SVP basis set
 requires 774 cluster amplitudes, and, therefore, several thousands of quantum gates. The unitary selective coupled cluster (USCC) method of Fedorov \textit{et al.}\cite{Fedorov2022} employs a selection scheme that includes only the most relevant excitations, using Hamiltonian matrix elements to identify the most connected excited state determinants to the single determinant ground state wave function. 

In this study, we generalize this approach to LAS-UCCSD. We utilize the Baker-Campbell-Hausdorff (BCH) expansion to re-express the USCC selection criterion in terms of LAS-UCCSD energy gradients. These analytical gradients 
 are then used to select the most important cluster amplitudes in a new method that we call LAS-USCCSD. We demonstrate the advantage of LAS-USCCSD over LAS-UCCSD in terms of computational cost on several molecular systems.  

 It is important to also note some recent advancements in system-specific iterative ansatz design in quantum computing~\cite{Ryabinkin2018,Kottmann2021,Lang2020,Grimsley2019,Ryabinkin2020,Tang2021,Mehendale2023,Sim2021,Halder2024}. One of the leading algorithms in this domain is ADAPT-VQE~\cite{Grimsley2019}, which iteratively builds an ansatz by adding fermionic operators, based on the expensive evaluation of energy gradients with respect to the variational parameters of these operators.  ADAPT-VQE's method of selecting operators at each iteration is based on their impact on energy reduction, which leads to a more efficient ansatz generation compared to standard VQE. For a detailed comparison of the USCC ansatz and ADAPT-VQE ansatz, Fedorov et al.'s work\cite{Fedorov2022} provides insightful analysis. Similar to the USCC method~\cite{Fedorov2022}, the excitation selection criteria in LAS-USCCSD for reflect the `importance' of an excitation but starting from a multireference wave function. While this may not yield the most compact ansatz like ADAPT-VQE, it has the benefit of not requiring additional measurements on the quantum computer for determining important coefficients/parameters, hence potentially being more cost-effective in terms of measurements on the quantum hardware. Unlike ADAPT-VQE, LAS-USCCSD is a single shot fixed-ansatz scheme \textit{i.e.} it determines the optimal amplitudes as a form of classical pre-screening and does not require any VQE/QPE calculations on the quantum computer for determining important coefficients/parameter.

 The paper is organized as follows. We first describe the LAS-UCCSD and USCC methods followed by our implementation of the LAS-USCCSD method. Next, we discuss numerical results obtained using LAS-USCCSD on four strongly correlated systems with important inter-fragment correlations. Finally, we provide an outlook of the method and discuss future possibilities.

\section{Theory and Methods}
\subsection{Localized Active Space Unitary Coupled Cluster Method}
In the LAS-UCCSD algorithm\cite{Otten2022} the energy is obtained by first performing a LASSCF calculation followed by a VQE on the combined fragment space to restore inter-fragment electron correlation~\cite{Otten2022,Dcunha2023StatePreparation}. LAS-UCCSD has been validated using Qiskit's Aer state vector simulator~\cite{Dcunha2023StatePreparation,Qiskit} and is yet to be tested on quantum hardware. 
The LASSCF wave function\cite{LASSCF0,LASSCF1} is an anti-symmetrized product of the \( K \)-fragment CAS wave functions and is represented as:
\begin{equation}
\ket{\Psi_{\text{LAS}}} = \bigwedge_{K} (\Psi_{A_K}) \wedge \Phi_D
\end{equation}

where \( \Psi_{A_k} \) denotes the many-body wave function of the \( K \)th localized subspace, and \( \Phi_D \) denotes the single determinantal wave function delocalized over the system under consideration.
Considering  $\ket{\Psi_{\text{QLAS}}} = \bigwedge\limits_{K} (\Psi_{A_K})$ as the active-space LASSCF wave function loaded onto a quantum device, the LAS-UCCSD wave function can be expressed as:
\begin{equation}
    |\Psi_{\textrm{LAS-UCCSD}} (\bm{t})\rangle  = \hat{U}_{\textrm{UCCSD}} (\bm{t})|\Psi_{\text{QLAS}} \rangle .
\end{equation}
Here, the corresponding VQE on the loaded LASSCF wave function is done using a generalized form of the UCC singles and doubles (UCCSD) ansatz~\cite{Lee2018,Hoffmann1988,McArdle2020}:
\begin{equation}
\hat{U}_{\textrm{UCCSD}} = \exp \left\{ t_{l}^k(\hat{a}_k^\dagger \hat{a}_l - \text{h.c.}) \right. \left. + \frac{1}{4} t_{ln}^{km}(\hat{a}_k^\dagger \hat{a}_m^\dagger \hat{a}_n \hat{a}_l - \text{h.c.}) \right\},
\label{eq:UCC}
\end{equation}
  where \(t_{l}^k\) and \(t_{ln}^{km}\) are the cluster amplitudes for single and double excitations, respectively. \(\hat{a}_k^\dagger\) and \(\hat{a}_l\) are the creation and annihilation operators, acting on the molecular orbitals $k$ and $l$ respectively. ``h.c." stands for the Hermitian conjugate of the preceding terms. In practice, the process requires decomposing the single exponential of a sum of generators into a product of exponentials of individual generators. This can be done in a variety of ways, one of the most common of which is the Suzuki-Trotter decomposition~\cite{Ostmeyer2023,Hatano2005} used here, to the first order.
 The fragment Hamiltonians are transformed to the qubit representation using a fermion-to-spin transformation, such as the Jordan-Wigner transformation~\cite{Fradkin1989}. 
The minimization of the total energy via VQE is performed by variation of the $\bm{t}$ with
a classical computer and algorithm, with the energies being evaluated on a quantum simulator.

One of the challenges for the practical implementation of this method is the large number of UCCSD parameters ($\bm{t}$ amplitudes) required for the VQE energy optimization. For example, an energy LAS-UCCSD calculation for [Cr$_2$(OH)$_3$(NH$_3$)$_6$]$^{3+}$ requires 774 parameters - as can be calculated using equations 8 and 9 - and  2126 iterations. 
Each iteration refers to a numerical optimization step of  $|\Psi_{\textrm{LAS-UCCSD}} (\bm{t})\rangle$ using the Broyden–Fletcher–Goldfarb–Shanno (BFGS) algorithm~\cite{BROYDEN1970,Fletcher1970,Goldfarb1970,Shanno1970} or some other classical optimization algorithm. In a later section ``Resource Estimates" we show that this translates to 36,360 CNOT gates and 54,612 single qubit gates, which is impractical with the current or near-term quantum computing capabilities even using a state-vector simulator. To address this, we have developed a physically motivated scheme that focuses on only the most relevant parameters. 

\subsection{Unitary Selective Coupled Cluster Method}
The unitary selective coupled cluster (USCC) method of Fedorov \textit{et al.}\cite{Fedorov2022} starts from a Hartree-Fock reference wave function. In the initial step the most important amplitudes are selected based on the following criterion:
\begin{equation}
|H_{\beta \textbf{0}}| = |\bra{\Phi_{\beta}}\hat{H}\ket{\Phi_{\textbf{0}}}| \geq \epsilon.
\label{condition-1}
\end{equation}
Here, $H_{\beta \textbf{0}}$ represents the one- and two-body electronic Hamiltonian matrix element, connecting the determinant $\Phi_{\beta}$ to $\Phi_{\textbf{0}}$, and $\epsilon$ is a user-defined cut-off. The amplitudes $t_{\beta}$ corresponding to these excitations are included in the UCCSD ansatz generation as required in equation \ref{eq:UCC}. 

\subsection{Localized Active Space Unitary Selected Coupled Cluster Method}
A direct translation of USCC to the LASSCF Hamiltonian has the following form 
\begin{equation}
|H_{\text{LAS}_{\beta \textbf{0}}}| = |\bra{\Psi_{\text{LAS}_{\beta}}}\hat{H}\ket{\Psi_{\text{LAS}_{\textbf{0}}}}| \geq \epsilon.
\label{eq:condition-1-LAS}
\end{equation}
Here, $\Psi_{LAS_{\beta}}$ and $\Psi_{LAS_{\textbf{0}}}$ are both multiconfigurational LASSCF wave functions, the former generated by the action of a particular coupled-cluster generator $\beta$ on the latter. Unlike the single-reference case, these two states are not necessarily orthogonal\cite{Hermes2020} 
and their Hamiltonian matrix element (equation \ref{eq:condition-1-LAS}) is not reducible to a simple closed-form expression of a handful of 1- and 2-body Hamiltonian matrix elements. 

To address this, we recontextualize the USCC selection criterion in terms of the energy gradients using the Baker-Campbell-Hausdorff (BCH) expansion with respect to the cluster amplitudes:
\begin{equation}
|H_{\beta \textbf{0}}| = \left| \frac{\partial E_{\text{UCC}}}{2\partial t_{\beta}} \right|_{\bm{t} = 0} \geq \epsilon.
\label{USCC-cond-1}
\end{equation}
We have rewritten the Hamiltonian matrix elements in terms of the gradients of the UCC energy expression. Equation \eqref{USCC-cond-1} can now be used to select the Hamiltonian matrix elements $H_{\beta \textbf{0}}$ and therefore the corresponding  $t_{\beta}$ amplitudes (Derivation for equation \eqref{USCC-cond-1} in section 1 of the Supplementary Information (SI)). Now, we express the LAS Hamiltonian matrix elements of equation \eqref{eq:condition-1-LAS} in terms of energy gradients as in equation \eqref{USCC-cond-1} and replace
equation \eqref{eq:condition-1-LAS} with
\begin{equation}
\left| \frac{\partial E_{\text{LAS-UCCSD}}}{\partial t_{\beta}} \right|_{\bm{t} = 0} \geq \epsilon.
\label{eq:LAS_USCC-cond-1}
\end{equation}

The cluster amplitudes $t_{\beta}$ for a given $\epsilon$ are now included in the UCC ansatz for the VQE step and we refer to this method as LAS-USCCSD. 

The LAS-UCCSD gradients are evaluated before running any optimization steps, using only the information from the LASSCF wave function. 
The gradients are evaluated on a classical computer and selecting the most important parameters for VQE optimization is a form of classical preprocessing. We exclusively focus on direct initialization (DI), which has been shown to be more advantageous than using QPE to load the fragment wave functions for fragments requiring less than 20 qubits in terms of gate count and Trotter steps~\cite{Dcunha2023StatePreparation}. However, the insights derived here are applicable to all forms of state preparation. In both LAS-UCCSD and LAS-USCCSD, using DI, the CI vectors of the individual fragments are loaded onto the quantum circuit for each fragment using one and two-qubit gates, which has dimensions $N_{\text{frag,K}}$. Here $N_{\text{frag,K}}$ is the number of spin orbitals in the $K^{th}$ fragment's active space. This approach entangles the fragment qubits during state preparation. The steps to perform a LAS-USCCSD calculation are summarized in Algorithm \ref{alg:LASUCCSD}.

\begin{algorithm}[H]
\caption{Localized Active Space Unitary Selective Coupled Cluster}
\begin{algorithmic}[1]
\State Run a classical LASSCF calculation 
\State Use $\left| \frac{\partial E_{\text{LAS-UCCSD}}}{\partial t_{\beta}} \right|_{\bm{t} = 0} \geq \epsilon$ and a user-defined $\epsilon$ to generate the custom LAS-USCCSD ansatz with the most important parameters ($\bm{t}$ amplitudes)
\State Initialize the multiconfigurational LASSCF state either using localized QPE circuits or direct initialization (DI) and the LAS-USCCSD ansatz as in 2
\State Run VQE with the custom ansatz to compute energy
\State Lower the value of $\epsilon$ and repeat steps 1-3 to update amplitudes and check for energy convergence
\end{algorithmic}
\label{alg:LASUCCSD}
\end{algorithm} 

Here we discuss how to calculate the total number of parameters ($t_{\beta}$ amplitudes) for a given active space. This task is combinatorial and involves initially calculating the total possible number of singles and doubles amplitudes, followed by subtracting the number of possible singles and doubles amplitudes within the fragment subspaces. This calculation depends on the number of spatial orbitals and the underlying fragmentation. Calculating the number of singles amplitudes is relatively straightforward, with an example provided later for \( \mathrm{H}_2 \). The number of doubles amplitudes in a given active space can be found by using:
\begin{equation}
f(n) = 6\binom{\binom{n}{2}}{2} + 2(n + 1)\binom{n}{2}
\label{eq:LASUCC-excitations-1}
\end{equation}
where \( n \) is the number of spatial orbitals (half the number of spin orbitals), \( \binom{n}{2} \) is a combinatorial term representing the number of ways to choose 2 items from a set of 
n distinct items and \( \binom{\binom{n}{2}}{2} \) refers to the number of unique pairs of unique pairs of a set of \( n \) distinct items. For a LAS-UCCSD calculation in which all excitations must involve at least two fragments, one can subtract the number for each fragment's ``internal" excitations from the number for the whole system using:
\begin{equation}
f_{\text{LAS-UCCSD}}(n, n_K, n_L, \ldots, n_{N_{\text{frag}}}) = f(n) - \sum_{K}^{N_{\text{frag}}} f(n_K).
\label{eq:LASUCC-excitations-2}
\end{equation}
This is done because we focus only on inter-fragment excitations. Equations \ref{eq:LASUCC-excitations-1} and \ref{eq:LASUCC-excitations-2} can be used to find the total number of doubles amplitudes. 

The USCC method also included disconnected triple and quadruple terms generated using the previously computed singles and doubles cluster amplitudes. In future work, where electron correlation beyond singles and doubles is necessary, we may consider including disconnected and connected triple and quadruple terms. 

\subsection{Computational Methods}
All the LAS-UCCSD and LAS-USCCSD calculations in this work were performed using the LAS-USCC repository\cite{LAS-USCC} which includes modified versions of the mrh code\cite{mrh} and utilizes the electron integrals and quantum chemical solvers from PySCF (version 2.3).\cite{PySCF0, PySCF1} The current implementation for computing LAS-UCCSD gradients in Step 2 of algorithm \ref{alg:LASUCCSD} has exponential time and memory cost since it is designed to evaluate the gradient at any arbitrary vector of amplitudes $\boldsymbol{t}$. However, because LAS-USCC only requires the gradient evaluated at the origin, $\boldsymbol{t}=0$, it is possible to implement a memory-efficient, polynomial-time algorithm for these gradients using the generalized Wick's theorem,\cite{Kong2010} and this will be pursued in future work and is discussed further in Section S04 of the SI. All noise-free simulations of state preparation using DI and measurement circuits were carried out using the Qiskit framework and the Aer state vector simulator~\cite{Qiskit}. The basis sets and active spaces, and localization used for the representative examples are described as the examples are introduced below.

\section{Results and Discussion}
\begin{figure}[H]
    \centering
    \includegraphics[scale=0.4]{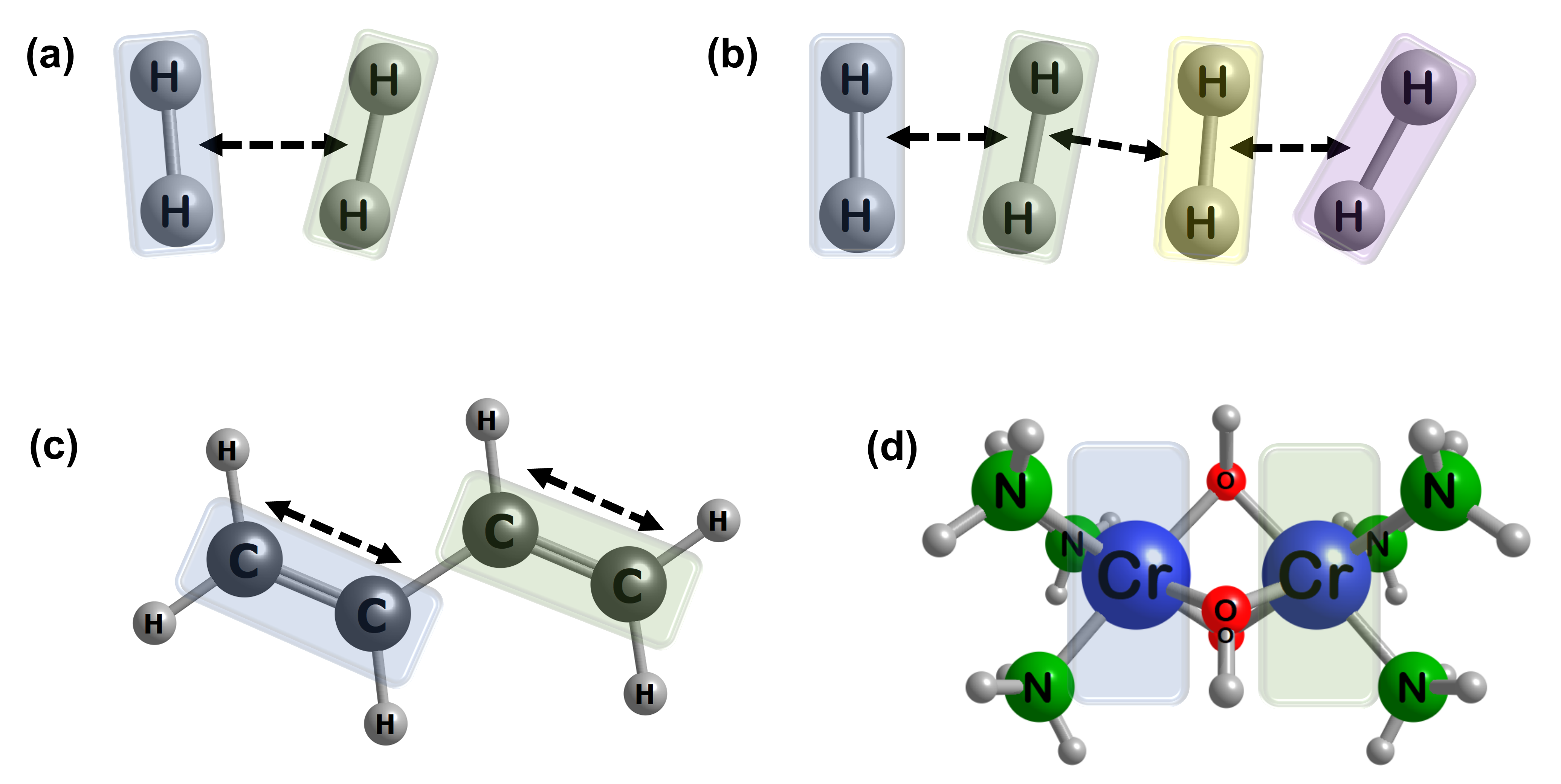}
    \caption{Systems studied in this work: (a) two
    interacting hydrogen molecules, (b) four
    interacting hydrogen molecules (c) the \textit{trans}-butadiene
    molecule, and (c) tris-(\(\mu\)-hydroxo)-bridged chromium molecule ([Cr$_2$(OH)$_3$(NH$_3$)$_6$]$^{3+}$) molecule. Arrows represent intermolecular and interatomic distances used to increase or decrease inter- and
    intra-fragment correlation for the hydrogen and \textit{trans}-butadiene systems respectively.}
    \label{fig:systems studied}
\end{figure}

Four distinct systems were considered, as illustrated in figure \ref{fig:systems studied}.
These systems present a different degree of strength of inter-fragment correlation, for which LASSCF shows significant deviations from the CASCI results and have been used as benchmark systems in references \citenum{Otten2022} and \citenum{Dcunha2023StatePreparation}.

\subsection{Hydrogen Molecules (\( \mathrm{H}_4 \) and \( \mathrm{H}_8 \))}

\begin{figure}[H]   
    \centering
    \includegraphics[scale=0.4]{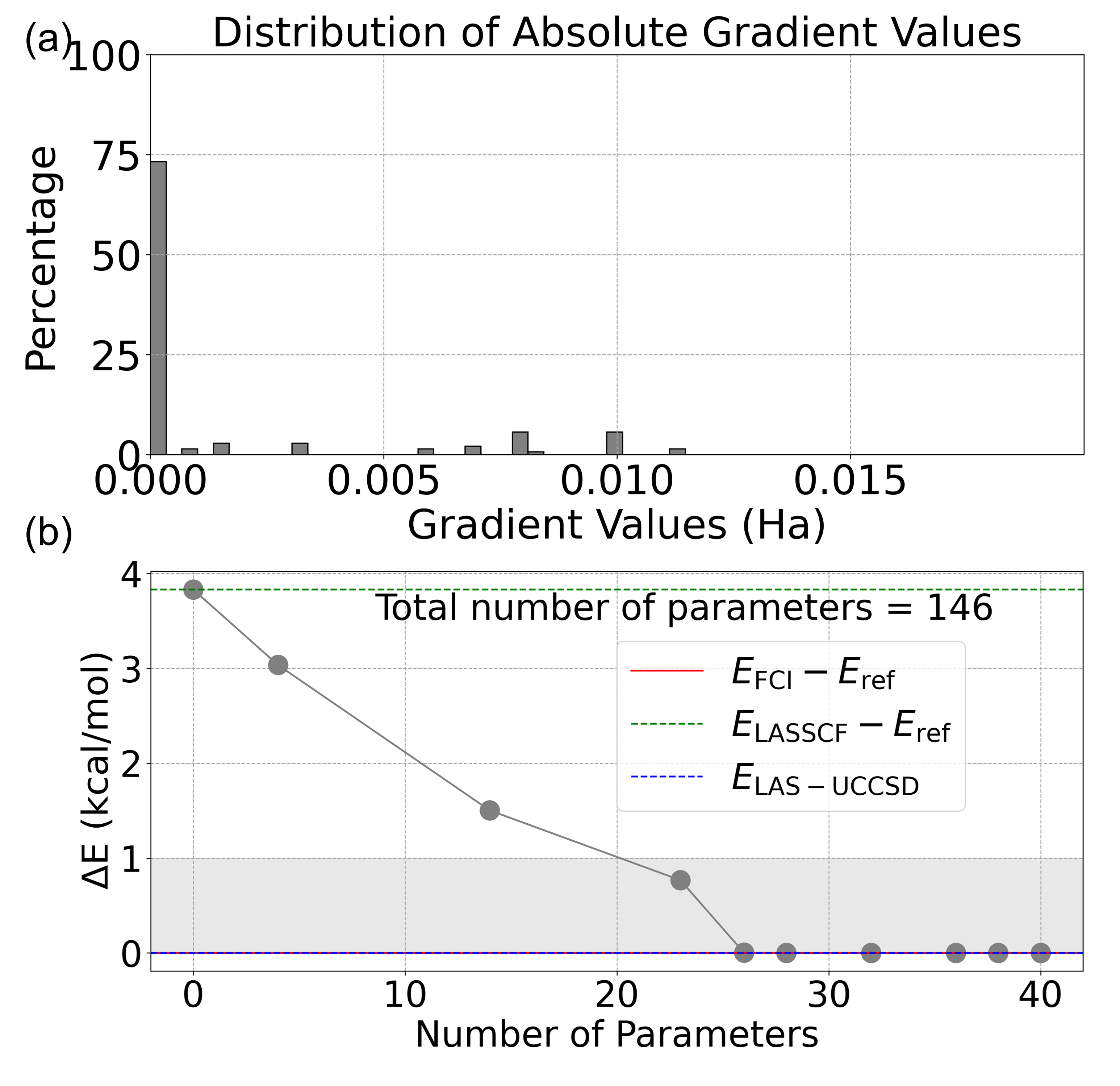}
    \caption{(a) Histogram displaying the percentage distribution of absolute gradient values (in $E_h$) for the H\(_4\) molecule at an inter-fragment distance of \(1.46 \, \text{\AA}\). (b) The energy convergence plot (\(\Delta E = E_{\text{LAS-USCCSD}} - E_{\text{LAS-UCCSD}}\)) for the LAS-USCCSD method as a function of the number of parameters, benchmarked against the LAS-UCCSD reference energy, showing the changes in calculated energy relative to the number of parameters included in LAS-USCCSD. The total number of parameters for the corresponding LAS-UCCSD calculation is 146. The shaded area highlights the region within 1 kcal/mol of the reference values. }
    \label{fig:H4-close}
\end{figure}

\begin{figure}   
    \centering
    \scalebox{0.5}{\includegraphics{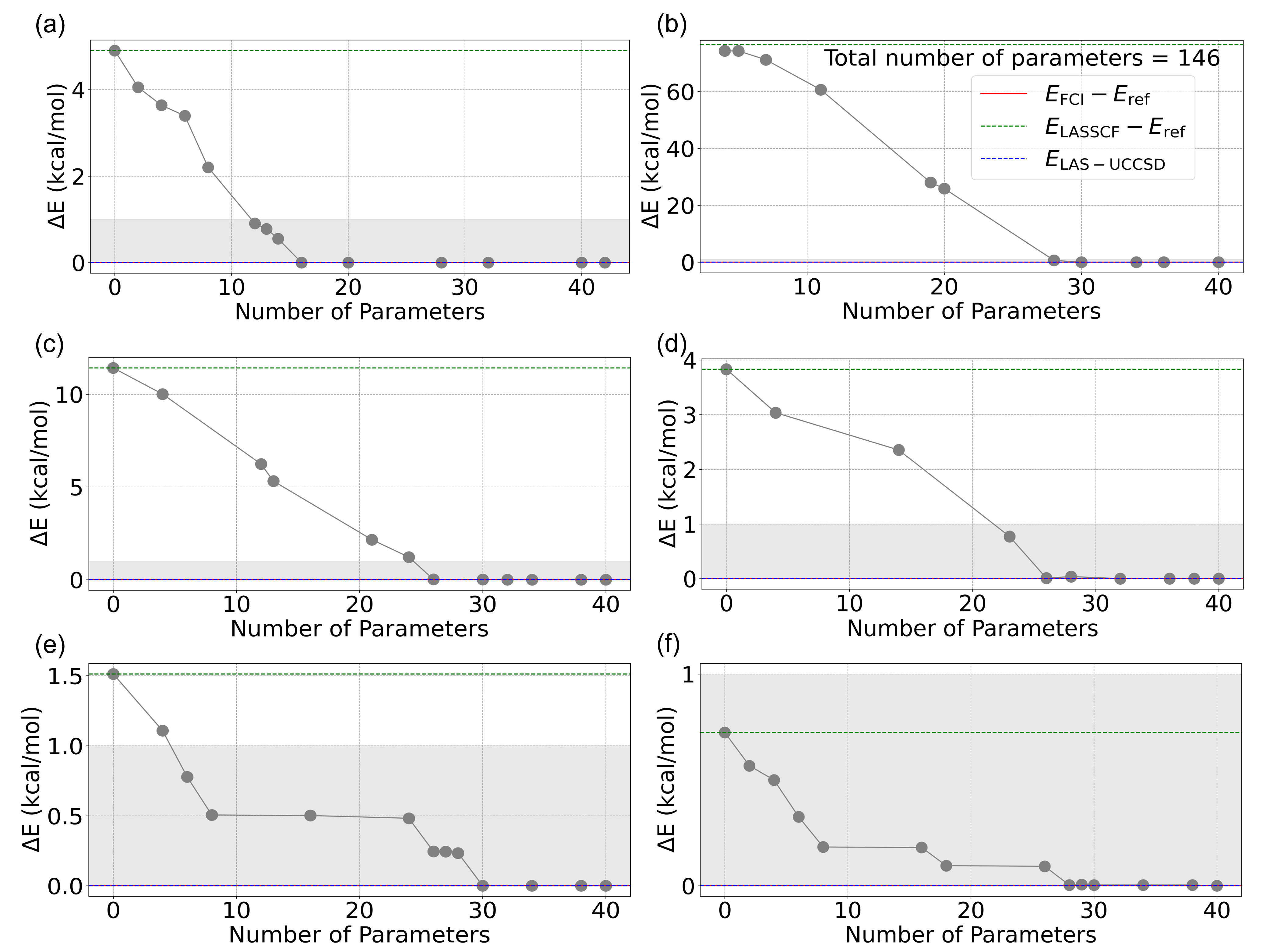}}
    \caption{Energy convergence (\(\Delta E = E_{\text{LAS-USCCSD}} - E_{\text{LAS-UCCSD}}\)) for the H\(_4\) molecule as a function of the number of parameters at varying inter-fragment distances. From top left to bottom right, the plots correspond to inter-fragment distances of (a) 0.75 \AA, (b) 0.96 \AA, (c) 1.20 \AA, (d) 1.46 \AA, (e) 1.75 \AA, and (f) 2.00 \AA. Each plot benchmarks the LAS-USCCSD energy against the LAS-UCCSD reference, illustrating how the energy difference decreases with an increasing number of parameters. The total number of parameters for all the corresponding LAS-UCCSD calculations is 146. The shaded area highlights the region within 1 kcal/mol of the reference values.}
    \label{fig:All-H4}
\end{figure}

The first system contains two \( \mathrm{H}_2 \) molecules with the distance between the midpoints of the two \( \mathrm{H}_2 \) fragments defining their interaction strength. We employed the STO-3G basis set and an active space of (2,2) for each fragment, corresponding to (4,4) in the CASCI calculations. The minimal STO-3G basis set was only used for the hydrogen systems for initial testing. Later, for the more realistic systems illustrated in Figure \ref{fig:systems studied}b-\ref{fig:systems studied}d, more complete basis sets were employed. The interaction is weak when the two \( \mathrm{H}_2 \) molecules are separated by more than \( 1.8 \, \text{\AA} \), and it becomes stronger as the two \( \mathrm{H}_2 \) molecules get closer~\cite{Otten2022}. We first focus on the two \( \mathrm{H}_2 \) molecules separated by 1.46 \( \text{\AA} \), where the total energy from LASSCF deviates from CASCI by \( 3.83 \, \text{kcal/mol} \). We calculated the gradient values according to equation \eqref{eq:LAS_USCC-cond-1} and plotted the distribution of number of parameters against the gradient values as a histogram in figure \ref{fig:H4-close}a. Using equations \ref{eq:LASUCC-excitations-1} and  \ref{eq:LASUCC-excitations-2} to obtain the number of doubles amplitudes, we obtain\( f(4) = 150 \) (the total number of spatial orbitals are 4), and \( f(2) = 6 \) (the fragments consist of 2 spatial orbitals each), from which \( f_{\text{LAS-UCCSD}} = 138 \). The number of singles amplitudes is 8 (since each electron has 2 choices), thereby leading to a total of 146 parameters (\(\bm{t}\) amplitudes). The data in figure \ref{fig:H4-close}a shows that around 70\% of the parameters have negligible gradient values, suggesting that they can be removed from the UCCSD ansatz. The dependence of the number of parameters on $\epsilon$ along with the total energies for this system is reported in table \ref{tab:H4-epsilon-vs-params} as obtained using equation \ref{eq:LAS_USCC-cond-1}. 
\begin{table}[H]
\centering
\begin{tabular}{ccc}
\toprule
$\epsilon$($E_h$) & Parameters & LAS-USCCSD Energy($E_h$)\\
\midrule
0.100000 & 0 & -2.102636 \\
0.062506 & 4 & -2.103902 \\
0.009541 & 14 & -2.106346 \\
0.007543 & 23 & -2.107519 \\
0.005964 & 26 & -2.108738 \\
0.004715 & 28 & -2.108741 \\
0.002947 & 32 & -2.108741 \\
0.001456 & 36 & -2.108741 \\
0.000910 & 38 & -2.108741 \\
0.000087 & 40 & -2.108741 \\
0.000000 & 146 & -2.108741 \\
\bottomrule
\end{tabular}
\caption{Dependence of the number of parameters on $\epsilon$ and the corresponding total energies for the $H_4$ system as is obtained using equation \ref{eq:LAS_USCC-cond-1}. The corresponding CASCI total energy is -2.108741 $E_h$.}
\label{tab:H4-epsilon-vs-params}
\end{table}
In figure \ref{fig:H4-close}a, we plot the distribution of absolute gradient values which indicates that a large percentage of gradients are near-zero. In figure \ref{fig:H4-close}b, we plot the energy discrepancy, denoted as $\Delta$E in kcal/mol, between the LAS-UCCSD and LAS-USCCSD methods. Both here and in table \ref{tab:H4-epsilon-vs-params}, we note that LAS-USCCSD achieves convergence to within 1 kcal/mol of the reference LAS-UCCSD energy with only 23 out of 146 parameters or 16\% of the parameter space.
For further insight, we also considered other distances between the two \( \mathrm{H}_2 \) molecules and plot the energy convergence with respect to the CASCI limit in figure \ref{fig:All-H4}. We observe that when the two \( \mathrm{H}_2 \) molecules are separated by \(0.96 \, \text{\AA} \), LASSCF diverges from LAS-UCCSD by more than 75 kcal/mol. This is because the distance between the two hydrogen atoms of a single  H$_2$ unit is \(1 \, \text{\AA} \) and this geometry is close to the equidistant square geometry. The discrepancy between LASSCF and LAS-UCCSD diminishes when the hydrogen moieties are brought even closer together (for example, when the intermolecular distance is \( 0.75 \, \text{\AA} \)) as the covalent bonds within each nominal H$_2$ molecule are broken and replaced with covalent bonds between the two nominal molecules, as described by the optimized LASSCF molecular orbitals, shown in figure S2 of the SI.  However, LAS-USCCSD with 28 parameters (an estimated 928 CNOT gates) out of 146 (6656 CNOT gates) restores back most of the electron correlation energy and is within 1 kcal/mol accuracy. The same behavior occurs for all the bond lengths studied. As was discussed in Otten \textit{et al.}~\cite{Otten2022}, LASSCF itself is enough to approximate the CASCI limit for inter-fragment distances greater than \(2.00 \, \text{\AA} \). 

\begin{figure}[H]   
    \centering
    \scalebox{0.4}{\includegraphics{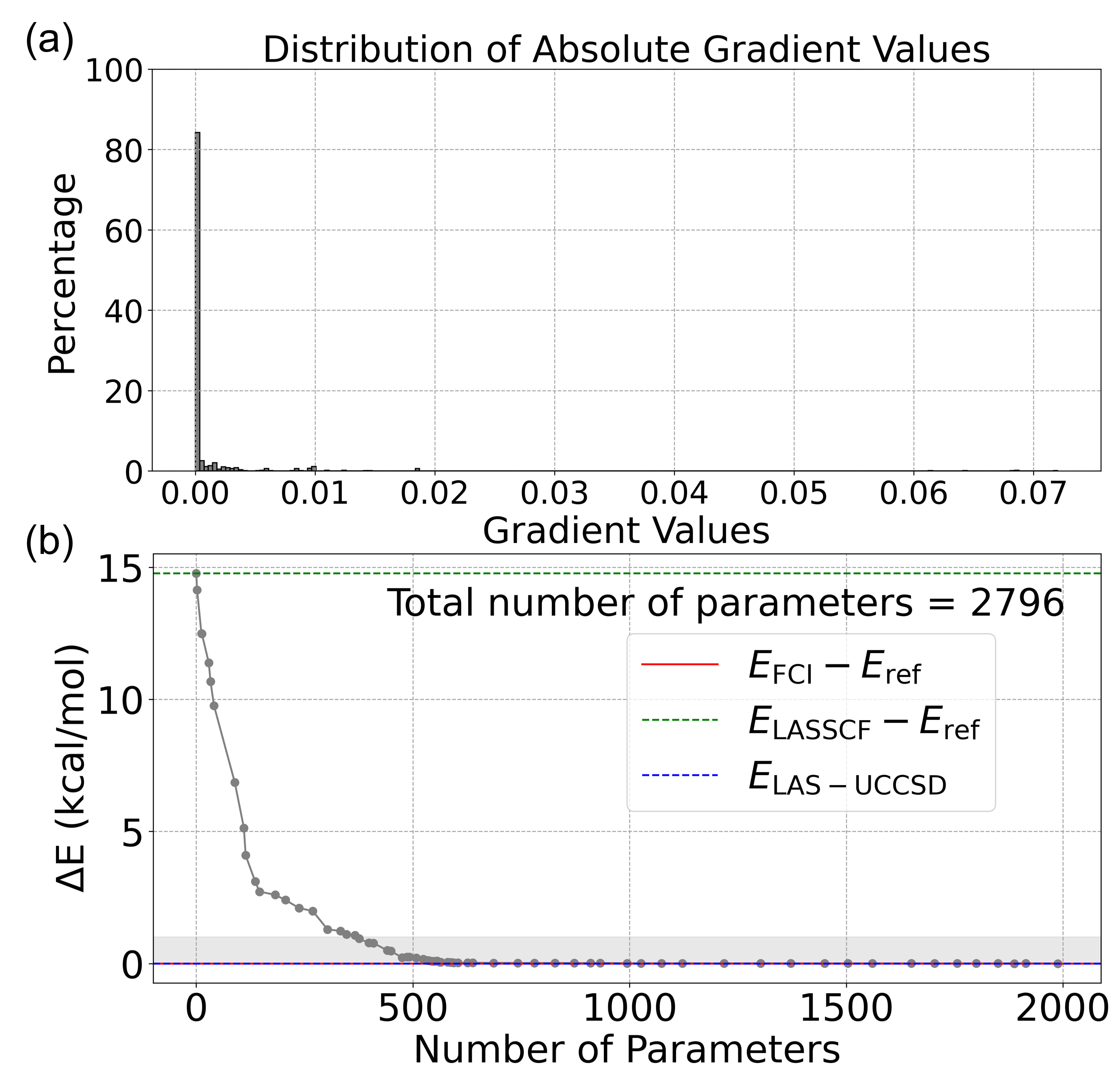}}
    \caption{(a) Histogram showing the percentage distribution of absolute gradient values (in $E_h$) for the H\(_8\) molecule at an inter-fragment distance of \( 1.46 \, \text{\AA}\). (b) Energy convergence plot (\(\Delta E = E_{\text{LAS-USCCSD}} - E_{\text{LAS-UCCSD}}\)) for H\(_8\), detailing the reduction in the calculated energy relative to the increasing number of parameters. The total number of parameters for the corresponding LAS-UCCSD calculation is 2796. The shaded area highlights the region of chemical accuracy defined as being within 1 kcal/mol of the reference values.}
    \label{fig:H8-close}
\end{figure}

Next, we study a system formed of four \( \mathrm{H}_2 \) molecules using a minimal STO-3G basis set for further testing in line with the previous example, with the distance between the midpoints of the two nearest neighbor fragments defining their interaction strength. We examine a geometry where the inter-fragment distance is \( 1.46 \, \text{\AA} \), a distance at which the \( \mathrm{H}_2 \) molecules are relatively close to one another, resulting in strong inter-fragment correlation. Under these conditions, LASSCF alone is insufficient to achieve chemical accuracy in calculating the total energies with respect to the corresponding CASCI. Similar to \( \mathrm{H}_4 \), in configurations where the four \( \mathrm{H}_2 \) molecules are farther apart, the LASSCF description suffices to reach chemical accuracy. We employed an active space of (2,2) for each fragment, corresponding to (8,8) in the CASCI calculations.
Here, LAS-UCCSD corresponds to 2796 parameters (equation \eqref{eq:LASUCC-excitations-2} can be used to get this number) to be optimized. We observe in figure \ref{fig:H8-close}a that for inter-fragment distances of \( 1.46 \, \text{\AA} \), around 80\% of the gradients are near zero, thereby suggesting that less than 20\% parameters are required for convergence. LASSCF diverges from LAS-UCCSD by almost 15 kcal/mol, whereas LAS-UCCSD using 398/2796 parameters ($\sim$15\% parameter space) \textit{i.e.} an estimated 27416/198336 CNOT gates, reaches chemical accuracy compared with LAS-UCCSD.

\subsection{\textit{Trans}-butadiene} 

\begin{figure}[H]   
    \centering
    \scalebox{0.4}{\includegraphics{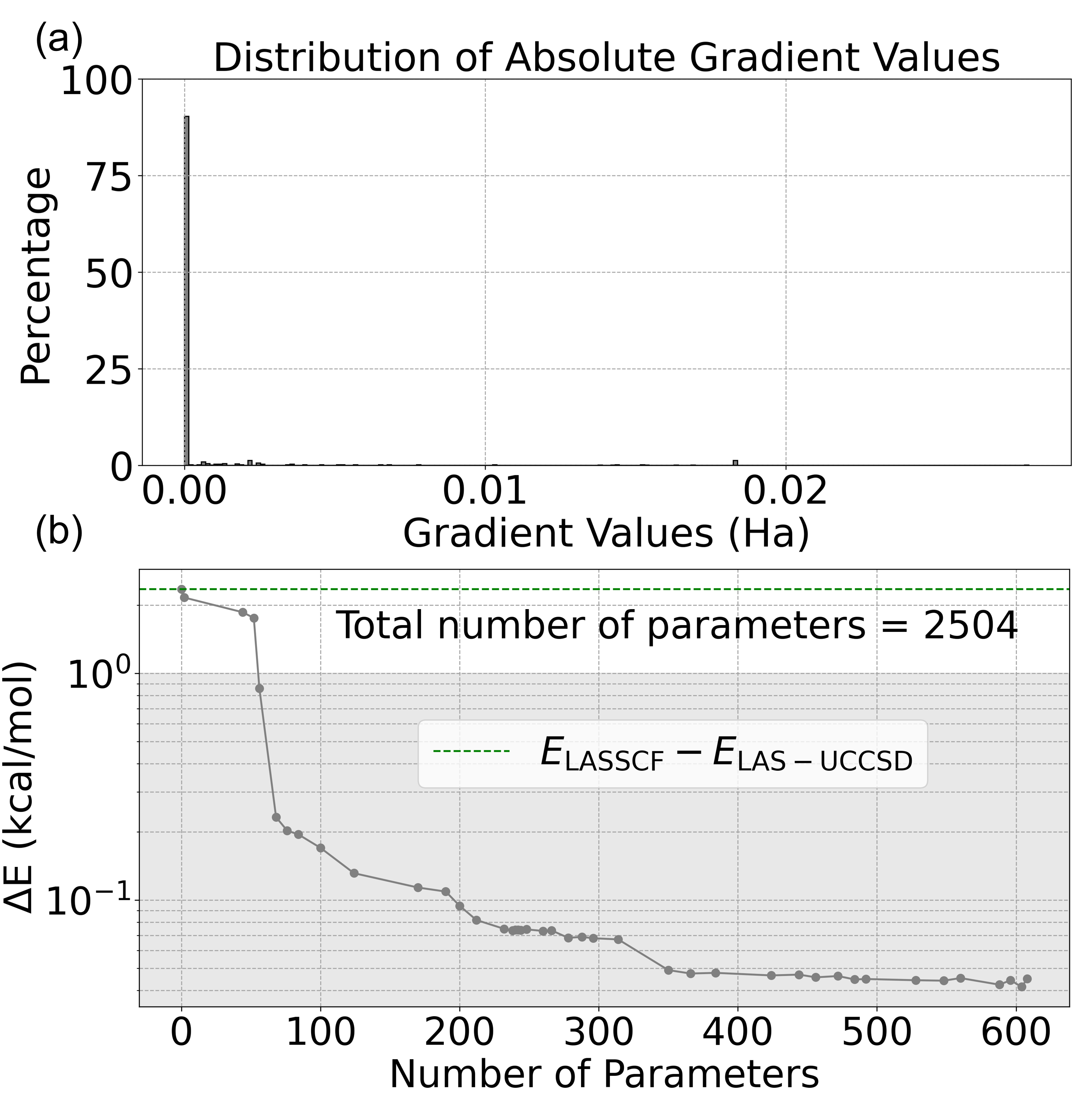}}
    \caption{(a) Distribution of absolute gradient values (in $E_h$) for geometry 1 of \textit{trans}-butadiene, showing the percentage of gradients within specified value ranges. (b) LAS-USCCSD energy convergence logarithmic plot (\(\Delta E = E_{\text{LAS-USCCSD}} - E_{\text{LAS-UCCSD}}\)) for stretched \textit{trans}-butadiene as a function of the number of parameters. The convergence is benchmarked against the LAS-UCCSD reference energy which has a total of 2504 parameters. The shaded area highlights the region within 1 kcal/mol of the reference values.}
    \label{fig:c4h6-equilibrium}
\end{figure}

\begin{figure}[H]   
    \centering
    \scalebox{0.4}{\includegraphics{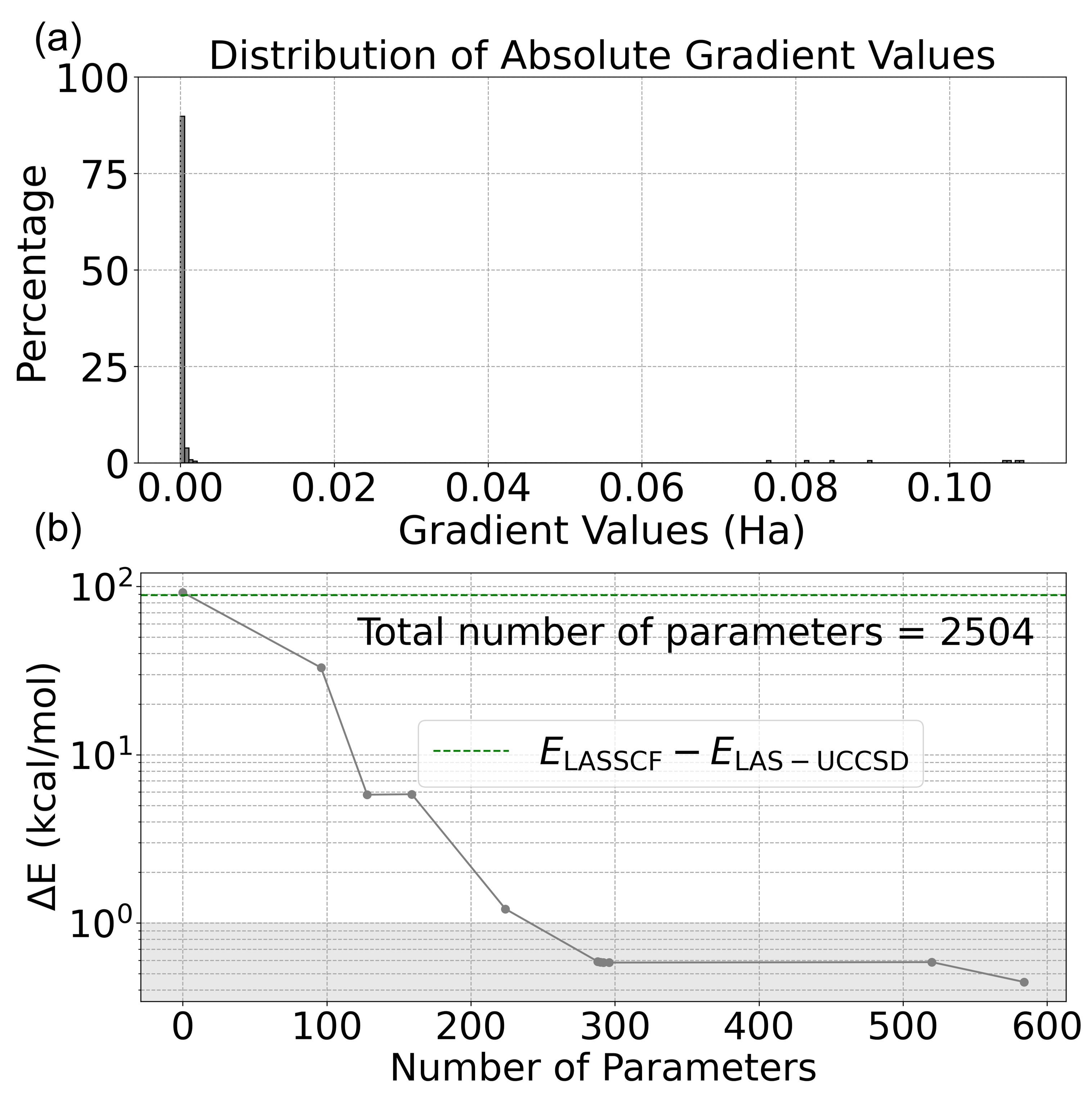}}
    \caption{(a) Distribution of absolute gradient values (in $E_h$) for geometry 2 of \textit{trans}-butadiene, showing the percentage of gradients within specified value ranges. (b) LAS-USCCSD energy convergence plot (\(\Delta E = E_{\text{LAS-USCCSD}} - E_{\text{LAS-UCCSD}}\)) for \textit{trans}-butadiene geometry 2 as a function of the number of parameters. The convergence is benchmarked against the LAS-UCCSD reference energy which has a total of 2504 parameters. The shaded area highlights the region of chemical accuracy defined as being within 1 kcal/mol of the reference values.}
    \label{fig:c4h6-stretched}
\end{figure}

We investigated \textit{trans}-butadiene using the 6-31G basis. It consists of two strongly interacting C-C double-bond fragments. At the equilibrium geometry (henceforth referred to as geometry 1) there is no significant inter-fragment electron correlation and the LASSCF energy is only 2.36 kcal/mol higher than the corresponding reference CASCI total energy. Figure \ref{fig:c4h6-equilibrium} shows that  ~90\% of the parameters can be removed for geometry 1 and in figure \ref{fig:c4h6-equilibrium}a we find that using only 84/2504 parameters (3856/118784 CNOT gates) we get to within chemical accuracy of the reference. The total number of parameters can be calculated using equations \ref{eq:LASUCC-excitations-1} and \ref{eq:LASUCC-excitations-2} and the number of parameters required to achieve chemical accuracy can be achieved using equation \ref{eq:LAS_USCC-cond-1} and $\epsilon = 4.14*10^{-3}$. The dependence of the number of parameters on $\epsilon$ are further reported in Table S5 of the SI. The CNOT gate estimates are discussed in the section ``Resource Estimates".
Let us now consider ``geometry 2" where the C-C double bonds are elongated by 3 $\text{\AA}$ from the equilibrium bond length of 1.33 $\text{\AA}$ while keeping the C-C single bond length fixed, thereby leading to a C-C double bond length of 4.33 $\text{\AA}$ (figure \ref{fig:systems studied}c). Figure \ref{fig:c4h6-stretched}a shows that ~90\% of the parameters have zero gradient value, implying that 288/2504 parameters are enough to achieve chemical accuracy, (figure \ref{fig:c4h6-stretched}b). Of note is that the gradients for geometry 2 (figure \ref{fig:c4h6-stretched}a) span up to an absolute value of 0.1 $E_h$ whereas the gradients for the equilibrium geometry span only up to 0.02 $E_h$ (figure \ref{fig:c4h6-equilibrium}a). For the particular case of geometry 2, the LAS-USCCSD with zero parameter energy is different from the LASSCF energy since LAS-USCCSD (0 parameters) gets initialized at a different local minimum and is therefore different from the LASSCF ground state. 
\FloatBarrier

\FloatBarrier
\subsection{Tris-(\(\mu\)-hydroxo)-bridged Chromium Compound}
\begin{figure}[H]   
    \centering
    \includegraphics[scale=0.4]{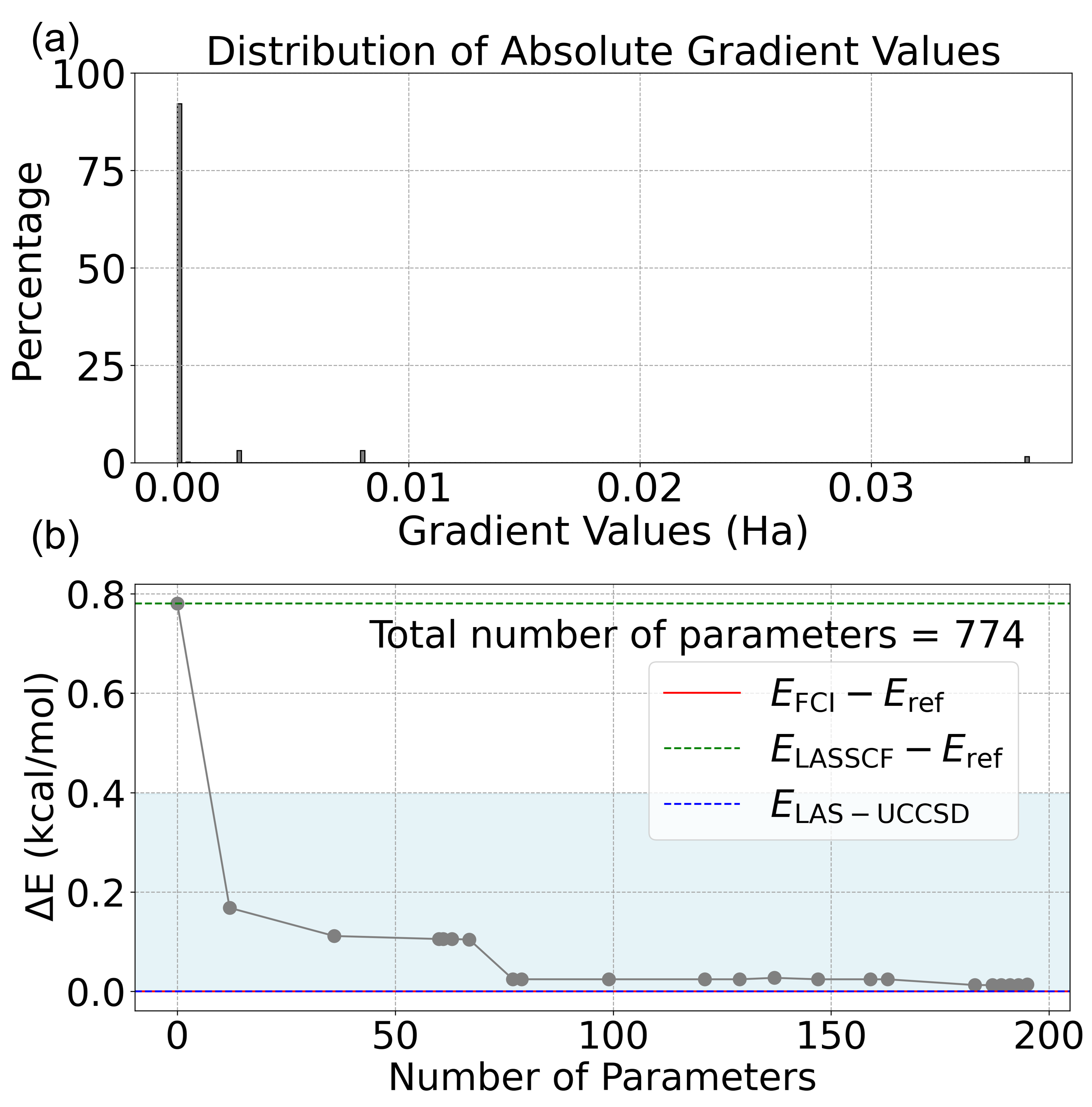}
    \caption{(a) The distribution of absolute gradient values (in $E_h$) for the chromium dimer system, indicating the percentage of total gradients within specific value ranges. (b) The convergence of energy (\(\Delta E = E_{\text{LAS-USCCSD}} - E_{\text{LAS-UCCSD}}\)) as a function of the number of parameters employed in the calculations, with the total number of parameters for LAS-UCCSD being 774. The reference energy is taken from the LAS-UCCSD method, and the graph shows how the energy difference decreases with an increasing number of parameters used. The shaded area in blue highlights the region of where the correct ground state is predicted and the low-spin state is lower than the high-spin state}
    \label{fig:chromium-dimer-LS}
\end{figure}

Finally, we compute  the magnetic coupling constant ($J$-coupling parameter) for the bimetallic compound [Cr$_2$(OH)$_3$(NH$_3$)$_6$]$^{3+}$, (figure \ref{fig:systems studied}d) with LAS-USCCSD. The $J$ value can be calculated using the Yamaguchi formula\cite{Yamaguchi1986} as the difference between the high-spin (\text{HS}) and low-spin (\text{LS}) energies divided by the difference between the $\left\langle \hat{S}^2 \right\rangle$ (expectation value of the $\hat{S}^2$ spin operator) values of the HS and LS states.
\begin{equation}
J_{ab} = \frac{E_{\text{HS}} - E_{\text{LS}}}{\left\langle \hat{S}^2 \right\rangle_{\text{LS}} - \left\langle \hat{S}^2 \right\rangle_{\text{HS}}}
\end{equation}
 A negative $J$-coupling indicates antiferromagnetic coupling, whereas
a positive $J$-coupling indicates ferromagnetic coupling. Otten et al. \cite{Otten2022} showed that LAS-UCCSD using a minimal active space of 6 electrons in 6 singly-occupied 3d orbitals, 3 for each Cr ion, predicts a $J$ value of -11.6 cm$^{-1}$ in agreement with the corresponding CASCI value of -11.6 cm$^{-1}$. The def2-SVP
basis was used for the C, N, O, and H atoms, while the def2-TZVP basis\cite{Weigend2005} was used for the Cr atoms. A LAS-UCCSD calculation requires 774 parameters (($\bm{t}$ amplitudes) and 2126 UCC iterations for convergence as also reported in figure \ref{fig:chromium-dimer-LS} and Table \ref{tab:Cr-dimer-classical}. However, ~90\% of the amplitudes have near-zero gradients as shown in figure \ref{fig:chromium-dimer-LS}a. LAS-USCCSD with only 12/774 parameters and 18 iterations predicts the correct sign for the $J$-coupling constant and the correct low-spin ground state. LAS-USCCSD with only ~10\% of the parameters is within 1 cm$^{-1}$ of the desired CASCI $J$-coupling value. On a state vector simulator\cite{Qiskit} with 12 qubits and first-order Trotterization, this calculation using direct initialization did not converge with our current computing resources. The LAS-UCCSD was run for 240 hours on a 256 Gb of memory on an AMD EPYC-7702 128-core processor with the VQE step reaching only the forty-sixth iteration with the first iteration taking up more than 28 hours. Using 12 parameters (488 CNOT gates), LAS-USCCSD converged within 73 seconds and using 77 parameters (3256 CNOT gates) it required less than 5 hours of computing time. For LAS-USCCSD using 12 parameters, the first VQE iterations take around 15 seconds and for 77 parameters, the first VQE iteration takes 7 minutes. Using the quantum circuit simulator with DI, LAS-USCCSD with 12 UCC parameters predicts the right ground state and LAS-USCCSD with 77 UCC parameters is within 1 cm$^{-1}$ of the desired CASCI $J$-coupling value.
\begin{table}[H]
\centering
\caption{Comparison of (\( J \)) values for the [Cr$_2$(OH)$_3$(NH$_3$)$_6$]$^{3+}$ across various computational methods used on a classical emulator, focusing on parameter counts and iteration numbers for convergence. The high spin is single configurational and therefore is the same as -2649.1455510 $E_h$ for all methods.}
\begin{tabular}{lcccc}
\hline
\textbf{Method} & \textbf{Parameters} & \textbf{Iterations} & \textbf{Low Spin Energy} & \textbf{J (cm\(^{-1}\))} \\
\hline
CASSCF & - & - & -2649.1463079 & -13.8 \\
LASSCF & - & - & -2649.1449395 & 11.2 \\
CASCI@LAS orbitals & - & - & -2649.1461837 & -11.6 \\
LAS-UCCSD & 774 & 2126 & -2649.1461837 & -11.6 \\
LAS-USCCSD & 12 & 18 & -2649.1459158 & -6.7 \\
LAS-USCCSD & 36 & 28 & -2649.1460061 & -8.3 \\
LAS-USCCSD & 77 & 169 & -2649.1461448 & -10.9 \\
LAS-USCCSD & 187 & 1980 & -2649.1461634 & -11.2 \\
\hline
\end{tabular}
\label{tab:Cr-dimer-classical}
\end{table}
\begin{table}[H]
\centering
\begin{threeparttable}
\caption{Comparison of (\( J \)) values for the [Cr$_2$(OH)$_3$(NH$_3$)$_6$]$^{3+}$ across various computational methods used on the state vector simulator, focusing on VQE time required for convergence. The high spin is single configurational and therefore is the same as -2649.1455510 $E_h$ for all methods.}
\begin{tabular}{lcccc}
\hline
\textbf{Method} & \textbf{Parameters} & \textbf{VQE Time} & \textbf{Low Spin Energy} & \textbf{J (cm\(^{-1}\))} \\
\hline
CASSCF & - & - & -2649.1463079 & -13.8 \\
LASSCF & - & - & -2649.1449395 & 11.2 \\
CASCI@LAS orbitals & - & - & -2649.1461837 & -11.6 \\
LAS-UCCSD & 774 & 240** & NA** & NA \\
LAS-USCCSD & 12 & 0.02 & -2649.1459158 & -6.7 \\
LAS-USCCSD & 77 & 4.43 & -2649.1461428 & -10.8 \\
\hline
\end{tabular}
\begin{tablenotes}
\small
\item ** These calculations have not converged.
\end{tablenotes}
\end{threeparttable}
\label{tab:Cr-dimer-state-vector}
\end{table}
\FloatBarrier
\section{Resource Estimates}
The number of single qubit gates ($\text{N}_{\text{SQGs}}$) and CNOT gates ($\text{N}_{\text{CNOTs}}$) calculated using the analytical formulas provided in reference \citenum{Magoulas2023} for standard qubit gate counts are provided in table \ref{table:gate_counts}. These can be written as:
\begin{subequations}
\begin{align}
\text{N}_{\text{SQGs}} &= (4n+1)2^{2n-1} \\
\text{N}_{\text{CNOTs}} &= (2n-1)2^{2n} 
\end{align}
As evidenced by the reduction in the number of parameters ($\bm{t}$ amplitudes), we also observe a dramatic reduction in the corresponding single qubit and CNOT gate counts. Less than 15\% of the total number of gates required in LAS-UCCSD are sufficient to obtain total energies within chemical accuracy of the corresponding CASCI value. For the [Cr$_2$(OH)$_3$(NH$_3$)$_6$]$^{3+}$ compound only 1.34\% of CNOT gates are required to predict the correct ground state and about  9\% of CNOT gates are important to predict the (\( J \)) values within 1 cm$^{-1}$ accuracy of the corresponding CASCI value.
\end{subequations}
\begin{table}[H]
\centering
\caption{Single qubit Gates (SQGs) and CNOT gates for LASUCCSD and LASUSCC for each system studied. The final column indicates the percentage of CNOT gates required by LASUSCCSD in comparison to their LASUCCSD counterparts. }
\begin{tabular}{llcccc}
\hline
System                       & Method     & Parameters & $\text{N}_{\text{SQGs}}$ & $\text{N}_{\text{CNOTs}}$ & \% $\text{N}_{\text{CNOTs}}$ \\
\hline
\( \text{H}_4 \)                     & LASUCCSD   & 146        & 10016              & 6656       & -\\
\( \text{H}_4 \)                     & LASUSCCSD  & 23         & 1408               & 928        & 13.94 \\
\( \text{H}_8 \)                     & LASUCCSD   & 2796       & 198336             & 132096     & - \\
\( \text{H}_8 \)                     & LASUSCCSD  & 398        & 27416              & 18224      & 13.80 \\
\( \text{C}_4\text{H}_6 \) geometry 1 & LASUCCSD   & 2504       & 178304             & 118784     & - \\
\( \text{C}_4\text{H}_6 \) geometry 1 & LASUSCCSD  & 84         & 5800               & 3856       & 3.46 \\
\( \text{C}_4\text{H}_6 \) geometry 2 & LASUCCSD   & 2504       & 178304             & 118784     & - \\
\( \text{C}_4\text{H}_6 \) geometry 2 & LASUSCCSD  & 288        & 19744              & 13120      & 11.05 \\
\( \text{[Cr}_2(\text{OH})_2(\text{NH}_3)_6\text{]}^{3+} \) & LASUCCSD   & 774        & 54612              & 36360      & - \\
\( \text{[Cr}_2(\text{OH})_2(\text{NH}_3)_6\text{]}^{3+} \) & LASUSCCSD  & 12         & 740                & 488        & 1.34 \\
\( \text{[Cr}_2(\text{OH})_2(\text{NH}_3)_6\text{]}^{3+} \) & LASUSCCSD  & 77         & 4924               & 3256       & 8.95 \\
\hline  
\end{tabular}
\label{table:gate_counts}
\end{table}

\section{Summary and Future Directions}

We have introduced an efficient implementation of a multireference quantum algorithm called localized active space unitary selective coupled cluster (LAS-USCCSD). It is derived from the parent classical LAS-UCCSD method. This algorithm considers only the most important parameters required to optimize a multireference wave function on a quantum circuit simulator, and although only validated on classical computers using the Qiskit statevector simulator, it is designed with early fault-tolerant quantum devices. We have tested the method on
systems ranging from H$_4$, H$_8$ with minimal basis sets to a bimetallic chromium dimer compound using a mixture of more complete def2-TZVP/def2-SVP basis sets. Our findings show an improvement in algorithm efficiency and faster processing times for LAS-USCCSD with respect to LAS-UCCSD, with a reduction of approximately 85\% to 90\% in the number of CNOT gates required to predict total energies in selected benchmark systems within chemical accuracy. Additionally, we observed a reduction of around 90\% in the number of parameters and CNOT gates for predicting the magnetic coupling constant in a bimetallic compound, [Cr$_2$(OH)$_3$(NH$_3$)$_6$]$^{3+}$ within 1 cm$^{-1}$ of the corresponding CASCI value. 
A future direction is to utilize generalized Wick's theorem for multireference wave functions to develop a polynomial-time algorithm for LAS-UCCSD gradient computation to increase memory efficiency for enhanced scalability. 
Furthermore, utilizing preselected LAS-USCCSD parameters in conjunction with adaptive and hardware-efficient quantum computing ansätze, such as ADAPT-VQE and nu-VQE, alongside efforts aimed at reducing CNOT gate counts~\cite{Burton2023, Magoulas2023} will lead to a further reduction in gate counts and circuit depth while maintaining accuracy. LAS-USCCSD will complement these methods for inherently multireference systems, potentially leading to more effective quantum simulations of realistic chemical systems and enabling the modeling of larger active spaces in the near future. Additionally, benchmarking the performance of ADAPT-VQE against both USCC and LAS-USCCSD approaches for inherently single and multi-configurational systems respectively is important and will be done in the future. Right now, classical calculations are still far superior, but with advancements in both quantum hardware and software, quantum computers may eventually overcome existing classical bottlenecks in quantum chemistry. 

\section*{Associated content}
\textbf{Supporting Information Available}. Derivation of USCC condition; Convergence of LAS-USCCSD for [Cr$_2$(OH)$_3$(NH$_3$)$_6$]$^{3+}$; Total Energies; LASSCF energy evaluation; LASSCF orbitals for H\(_4\)

 This material is available free of charge via the Internet at http://pubs.acs.org. \\

 \section*{Author Information}
\textbf{Corresponding Authors} \\
*lgagliardi@uchicago.edu (L.G.), *gray@anl.gov (S.G.),  *mjotten@wisc.edu (M.O.)\\
\textbf{Author Contributions} \\
The manuscript was written through contributions of all authors. All authors have given approval to the final version of the manuscript. \\
\textbf{Notes} \\
The authors declare no competing financial interests.
\acknowledgement
We thank Kanav Setia and Kenny Heitritter for their insightful discussion. This material is based upon work supported by the U.S. Department of Energy, Office of Science, National Quantum Information Science Research Centers. We gratefully acknowledge support from the NSF QuBBE Quantum Leap Challenge Institute (NSF OMA-2121044). Computer resources were provided by the Research Computing Center at the University of Chicago. Y.A. and S.G. acknowledge support from the U.S. Department of Energy, Ofﬁce of Science, under contract DE-AC02-06CH11357 at Argonne National Laboratory. Part of the work is supported by Wellcome Leap as part of the Quantum for Bio Program. Part of this work was also supported by Laboratory Directed Research and Development (LDRD) funding from  Argonne National Laboratory, provided by the Director, Office of Science, of the U.S. DOE under Contract no. DE-AC02-06CH11357. Work performed at the Center for Nanoscale Materials, a U.S. Department of Energy Office of Science User Facility, was supported by the U.S. DOE, Office of Basic Energy Sciences, under Contract No. DE-AC02-06CH11357.
\newpage

\FloatBarrier
\section*{ToC Graphic}
\begin{figure}[ht]
	\centering
\includegraphics[width =3.25in]{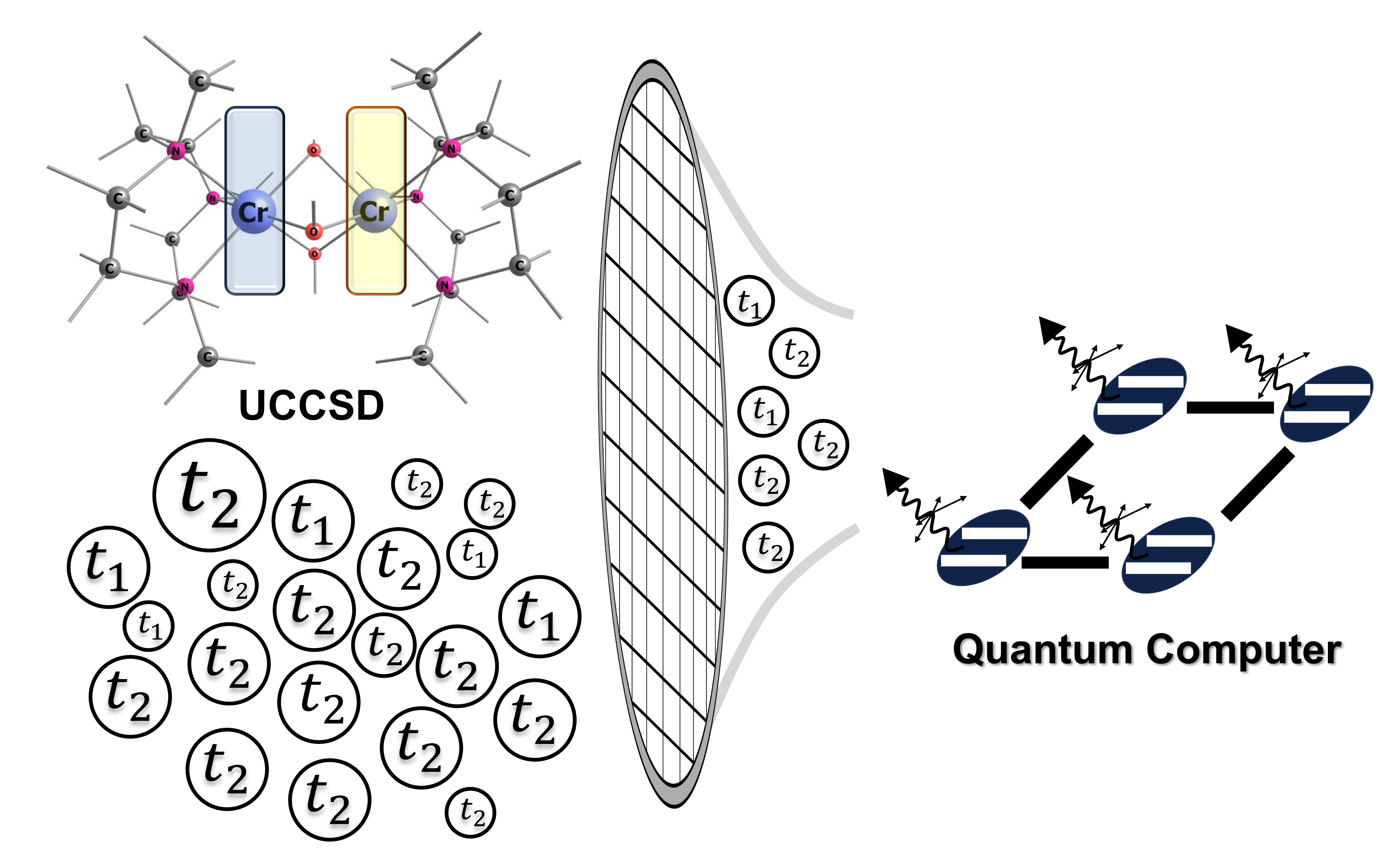}
\end{figure}
\FloatBarrier
\bibliography{achemso-demo}

\end{document}